\documentstyle[twocolumn,aps,psfig]{revtex}
\begin{document}             
\draft
\wideabs{          
\title{The onset of the vortex-like Nernst signal above $T_c$ in $\rm\bf La_{2-x}Sr_xCuO_4$ and $\rm\bf 
Bi_2Sr_{2-y}La_yCuO_6$.} 
\author{Yayu Wang$^1$, Z. A. Xu$^{1,\dagger}$, T. Kakeshita$^2$, S. Uchida$^2$, S. Ono$^3$, Yoichi 
Ando$^3$, and N. P. Ong$^1$}
\address{
$^1$Joseph Henry Laboratories of Physics, Princeton University, Princeton, New Jersey 08544, U.S.A..}
\address{
$^2$School of Frontier Sciences, University of Tokyo, Yayoi 2-11-16, Bunkyo-ku, Tokyo 113-8656, Japan. }
\address{
$^3$Central Research Institute of Electric Power
Industry (CRIEPI), Komae, Tokyo 201-8511, Japan}
\date{\today}      


\maketitle           

\begin{abstract}
The diffusion of vortices down a thermal gradient produces a Josephson signal which is detected as the vortex Nernst 
effect.  In a recent report, Xu {\it et al.}, Nature {\bf 406}, 486 (2000), an enhanced Nernst signal identified with 
vortex-like excitations was observed in a series of $\rm La_{2-x}Sr_xCuO_4$ (LSCO) crystals at temperatures 
50-100 K above $T_c$.  To pin down the onset temperature $T_{\nu}$ of the vortex-like signal in the lightly doped 
regime ($0.03 \le x \le 0.07$), we have re-analyzed in detail the carrier contribution to the Nernst signal.  By 
supplementing new Nernst measurements with thermopower and Hall-angle data, we isolate the off-diagonal Peltier 
conductivity $\alpha_{xy}$ and show that its profile provides an objective determination of $T_{\nu}$.  With the new 
results, we revise the phase diagram for the fluctuation regime in LSCO to accomodate the lightly doped regime.  In the 
cuprate $\rm Bi_2Sr_{2-y}La_yCuO_6$, we find that the carrier contribution is virtually negligible for $y$ in the range 
0.4-0.6. The evidence for an extended temperature interval with vortex-like excitations is even stronger in this system.  
Finally, we discuss how $T_{\nu}$ relates to the pseudogap temperature $T^*$ and the implications of strong 
fluctuations between the pseudogap state and the $d$-wave superconducting state.  

\end{abstract}
}				

\section{Introduction}\label{intro}
\begin{figure}[htb]
\centerline{\psfig{file=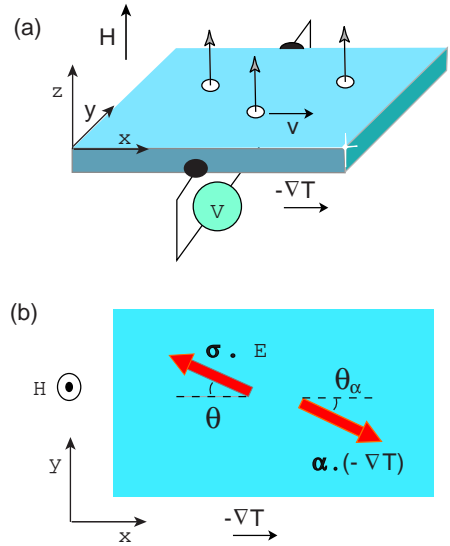,width=2.3in}}
\caption{(a) Geometry of the Nernst experiment in the vortex liquid state.  Vortices (disks with vectors) flow with 
velocity $\bf v$ down the gradient $-\nabla T\parallel {\bf \hat{x}}$.  Phase slippage induces a dc signal $E_y$ that is 
antisymmetric in $H$.  The lower panel (b) shows the currents $J_i = \sigma_{ij}E_j$ and $J'_i = \alpha_{ij}(-\partial_j 
T)$ produced by the $E$-field and thermal gradient, respectively, in the normal state.  The slight difference between their 
$y$-components engenders the carrier Nernst signal (angles $\theta$ and $\theta_{\alpha}$ are defined in the text).
}
\label{Expt}
\end{figure}

Close to the the upper critical field line $H_{c2}(T)$ of a type-II superconductor, the vortices exist in the `vortex liquid' 
state as highly mobile excitations.  In this mobile state, the vortices readily flow in response to a weak applied 
temperature gradient towards the cooler end of the sample.  By the Josephson effect, the vortex motion generates an 
electric field $\bf E = B\times v$ that lies perpendicular to both the vortex velocity $\bf v$ and ${\bf B} = \mu_0{\bf 
H}$ (Fig. \ref{Expt}(a)).  In general, the appearance of a transverse $\bf E$ in the presence of a thermal gradient and 
magnetic field is known as the Nernst effect \cite{Stephen}.  In weak fields, the Nernst {\em signal} $E_y/|\nabla T|$ 
increases linearly with $B$, but at higher fields, the curve of $E_y$ vs. $T$ tends to develop negative curvature.  The 
Nernst coefficient $\nu$ is defined as the Nernst signal per unit $B$ in the weak-$B$ limit.  Because a 
field-antisymmetric $E_y$ may be measured to high resolution, the Nernst effect provides a highly sensitive probe for 
detecting vortices.

From the purview of conventional superconductivity, the search for vortices high above the critical temperature $T_c$ 
seems quite unrewarding.  Above $T_c$, the average value of the condensate density $n_s$ is zero.  Although 
fluctuation effects produce small evanescent droplets of superconductivity detectable by susceptibility and resistivity, the 
existence of a Nernst signal in the fluctuation regime above $T_c$ is not expected.  Indeed, vortex Nernst signals in the 
fluctuation regime do not seem to have been reported for any low-$T_c$ superconductor.  In the first Ettinghausen 
\cite{Stephen} experiment on a cuprate ($\rm YBa_2Cu_3O_7$), however, Palstra {\it et al.} \cite{Palstra} noted that 
the signal extended  $\sim$10 K above $T_c$.   
\noindent
Later Nernst experiments on cuprates (restricted to optimally doped samples) found that the vortex signal extends above 
$T_c$ by roughly the same interval \cite{Hagen,Ri,Clayhold}.  Because the relative temperature interval $\sim 0.1$ is 
small, these observations did not provoke much theoretical comment.  More recently, Corson {\em et al.} \cite{Corson} 
investigated the conductivity in thin-film $\rm Bi_2Sr_2CaCu_2O_8$ at Terahertz frequencies and found that the kinetic 
inductance of the superfluid is observed as high as 25 K above $T_c$.  

In Nernst measurements on $\rm La_{2-x}Sr_xCuO_4$ (LSCO), Xu {\it et al.} \cite{Xu} observed that the Nernst 
signal remains anomalously enhanced 50-100 K above $T_c$, from which they inferred the existence of vortex-like 
excitations in the putative `normal-state'.  Starting at a relatively small and nearly $T$-independent value at high 
temperatures, $\nu$ begins to diverge at an onset temperature $T_{\nu}$, until it peaks near $T_c$ at a value 500 to 
1,000 times larger than the value at onset.  Because of the specific symmetry properties of the Nernst signal (${\bf E 
\parallel B\times }\nabla T$), Xu {\it et al.} argued that the excitations are vortex-like.   Moreover, they noted that the 
anomalous signal is very sensitive to Nd doping (which suppresses $T_c$ to $\sim$8 K), and {\em smoothly evolves to 
the familiar vortex signal} below $T_c$.  A closely similar extended fluctuation regime has also been observed in 
underdoped $\rm YBa_2Cu_3O_y$ \cite{Wang}.

The important issue whether the excitations are identical with magnetic vortices or are more exotic excitations (for e.g., 
holes bound to vorticity in a novel electronic state) is at present open.  Indeed, the notion that vortices of whatever origin 
can be detected 50-100 K {\em above} $T_c$ remains as surprising and challenging to us (and many in the community) 
as when they were first observed.  Hence, as in Ref. \cite{Xu}, we refer to them generically as {\em vortex-like} 
excitations in this paper.  We also refer to the region between $T_c$ and $T_{\nu}$ generically as the `fluctuation' 
regime.  (Discussion of this issue in relation to the pseudogap state is given in Sec. \ref{pseudo} .)

Our goal in this report is to sharpen the distinction of the vortex-like signal from that produced by normal-state charge 
carriers in order to address several issues raised by Ref. \cite{Xu} (we also provide extended discussions of many issues 
that were only touched on there).  The onset temperature of the vortex signal $T_{\nu}$ appears to increase with 
decreasing hole density $x$.  Is $T_{\nu}$ finite in the $x\rightarrow$ 0 limit?  If not, how does $T_{\nu}$ behave in 
this limit?  How significant is the normal-state carrier contribution to $\nu$ in the small-$x$ regime?  Can a more 
objective separation of the vortex signal be obtained through a better understanding of the normal-state Nernst effect?  
Are these excitations seen in other cuprates? 

The small-$x$ regime requires a more careful analysis of the carrier Nernst signal because the normal-state thermopower 
is strongly enhanced.  We introduce a new method suitable for this limit and apply it to LSCO and the single-layer 
cuprate $\rm Bi_2Sr_{2-y}La_yCuO_6$ (Bi 2201).  We isolate the key quantity in the Nernst experiment -- the 
off-diagonal Peltier current -- and show that it provides an objective procedure for separating the vortex signal that is 
particularly suited for the small-$x$ regime.  The new results enable us to obtain a revised phase diagram for fluctuations 
in LSCO that is valid in both the small- and large-$x$ regimes.  In Bi 2201, we find that the vortex-like signal, relative to 
the normal carrier contribution, is even larger than in LSCO.  Hence, $T_{\nu}$ may be obtained directly from the 
original Nernst signal without the need for isolating the Peltier current.

\section{Isolating the Peltier current}\label{isolate}
To explain our procedure, we recall the various charge currents generated in a Nernst experiment (initially we consider 
the normal-state terms only).  We take the thermal gradient $-\nabla T\parallel {\bf\hat{x}}$ and the field $\bf H\parallel 
{\bf \hat{z}}$.  The Nernst signal is the $H$-antisymmetric electric field $\bf E\parallel {\bf\hat{y}}$ per unit gradient 
(Fig. \ref{Expt}).  

In zero field, the applied gradient drives a charge current density ${\bf J} = \alpha(-\nabla T)$ along the length of the 
sample (we define the Peltier conductivity tensor $\alpha_{ij}$ by $J_i = \alpha_{ij}(-\partial_j T)$ and write for brevity 
$\alpha_{ii} = \alpha$).  To satisfy the boundary condition $J_x=0$, there must exist an $\bf E$-field to drive a current 
$\sigma E_x$ in the opposite direction ($\sigma$ is the electrical conductivity).  Hence, the total current along ${\bf 
\hat{x}}$ is
\begin{equation}
J_x = \sigma E_x + \alpha(-\partial_x T).
\label{Jx}
\end{equation}
With $J_x$= 0, we have $E_x = -(\alpha/\sigma)(-\partial_x T)$, which is the signal detected in a thermopower 
experiment.  The thermopower coefficient is $S = \alpha/\sigma$.  

In finite field, the two currents in Eq. \ref{Jx} spawn Hall-type currents (antisymmetric in $H$) flowing along the 
$y$-axis (Fig. \ref{Expt}(b)).  One is the familiar Hall current $\sigma_{yx}E_x$ while the other is the {\em 
off-diagonal} Peltier current $\alpha_{yx}(-\partial_x T)$, where $\sigma_{yx}$ and $\alpha_{yx}$ denote the Hall 
and off-diagonal Peltier conductivity, respectively.  These two off-diagonal currents are opposite in direction and nearly 
equal in magnitude.  (The standard Boltzmann-theory expressions for these currents are given in the Appendix.)  Because 
of the experimental boundary condition $J_y$ = 0, any residual difference between the off-diagonal currents leads to a 
weak $E_y$ which is then detected as the Nernst signal.  Hence, we have 
\begin{eqnarray}
J_y &= & \alpha_{yx}(-\partial_x T)  + \sigma_{yx}E_x + \sigma E_y \nonumber\\
& = & \left[\alpha_{yx} - \sigma_{yx}\frac{\alpha}{\sigma}\right](-\partial_x T) + \sigma E_y = 0.
\label{Jy}
\end{eqnarray}
We have dropped a term $\alpha(-\partial_y T)$, which is important in conventional metals, but negligible in cuprates 
(see Appendix).  Using the Hall angle $\tan\theta = \sigma_{xy}/\sigma$ in Eq. \ref{Jy}, we obtain for the Nernst 
coefficient due to charge carriers alone
\begin{equation}
\nu_N = \frac{E_y}{|\partial_x T|B} = 
\left[\frac{\alpha_{xy}}{\sigma} - S \tan\theta\right]\frac1B.
\label{nuN}
\end{equation}

At first glance, it may seem that $\nu_N$ is just the thermopower $S$ reduced by the Hall angle $\tan\theta$ (since it 
derives from a current transverse to the applied gradient).  However, by writing Eq. \ref {nuN} as $\nu = 
S[\tan\theta_{\alpha}-\tan\theta]/B$, with $\tan\theta_{\alpha}\equiv \alpha_{xy}/\alpha$, we see that the reduction 
factor involves a cancellation between the 2 angles $\theta$ and $\theta_{\alpha}$ (see Fig. \ref{Expt}(b)).  As shown 
by Sondheimer \cite{Sondheimer}, within Boltzmann theory the cancellation is exact if $\theta$ is independent of energy 
$\epsilon$ (see Appendix).  [Because of this cancellation, the ratio $E_y/E_x$ in a Nernst experiment does not 
represent a `Hall angle' for entropy currents.]

A useful order-of-magnitude estimate of $|\nu_N|$ is $|S\tan\theta/B|$ reduced by a factor of 10 (to account for the 
Sondheimer cancellation).  For LSCO in the range $0.1<x<0.17$, $S\simeq 10\;\mu$V/K and $\tan\theta/B \leq 
10^{-2}$ T$^{-1}$, we estimate $|\nu_N|\leq$ 10 nV/KT, which is what is generally observed.  This rule-of-thumb 
anticipates that when $S$ is of the order of $100\;\mu$V/K, $|\nu_N|$ may become as large as 50 nV/KT (see below).

{\em Diagonal and off-diagonal response}  Even if we disregard the Sondheimer cancellation, the carrier Nernst signal 
(being off-diagonal) is `small' compared with the (diagonal) thermopower signal whenever $\tan\theta\ll 1$.  By contrast, 
the vortex-Nernst signal represents the `large' response of the vortices to an applied $-\nabla T$.  If the flux-flow Hall 
angle $\tan\theta_f\ll 1$ (as in the cuprates), the vortex velocity $\bf v$ is very nearly $\parallel(-\nabla T)$.  Hence the 
Josephson field $\bf E_J$ represents the diagonal response of the flux motion (even though it appears as a transverse 
Hall-type signal).  On the other hand, the small velocity component $v_y=v_x\tan\theta$ transverse to $-\nabla T$ leads 
to a signal detected as the flux-flow  thermopower $S_f$, which {\em is the small off-diagonal signal} in the vortex 
liquid.  This reversal of roles for the Nernst and thermopower signals in going from the normal to the vortex-liquid state 
reflects the well-known duality between vortex and charge currents, and is the primary reason why the Nernst 
experiment is so useful for detecting vortex motion. The vortex Nernst signal reflects the {\em primary} response of 
vortices to an applied gradient, while the carrier Nernst effect is a relatively feeble off-diagonal response that is further 
attenuated by the Sondheimer cancellation.

For our present purpose, we are interested in the temperature range when the diagonal vortex signal has fallen to values 
comparable with the carrier signal.  The contribution of the vortices to the observed Nernst signal may be written as an 
off-diagonal term $\alpha^s_{xy}$ that adds to the normal-state term $\alpha^n_{xy}$ (now relabelled with `$n$').  
Repeating the steps above, we find that the combined Nernst signal is comprised of 3 terms, viz.
\begin{equation}
\nu = \frac{E_y}{|\partial_xT | B}  = 
\left[\frac{\alpha^s_{xy}}{\sigma} + 
\frac{\alpha^n_{xy}}{\sigma} - S \tan\theta\right]\frac1B. 			
\label{nuall}
\end{equation}

In Xu et al. \cite{Xu}, we assumed that the Sondheimer cancellation was sufficiently complete to reduce $\nu_N$ below 
a threshold of $\sim$4 nV/KT.  Hence, any increase of $\nu$ above this threshold was identified with the vortex term 
$\alpha^s_{xy}/\sigma$.  This `threshold criterion' is valid for moderately large $x$ where $|S|$ is small.  However, 
when $x\ll 1$, a new approach is needed.  For samples with $x\le$ 0.07, the thermopower $S$ (which sets the scale for 
$\alpha_{ij}$) rises to values in the range 100-300 $\mu$V/K.  This strong enhancement means that $\nu_N$ may 
significantly exceed the threshold, the Sondheimer cancellation notwithstanding.  

Our new method is based on measuring separately $S$ and $\tan\theta$ at each $T$.  By subtracting the product 
$-S\tan\theta$ from $\nu$ in Eq. \ref{nuall}, we obtain the total off-diagonal Peltier term 

\begin{equation}
\left[\frac{\alpha_{xy}}{\sigma}\right]_{obs} = \frac{\alpha^s_{xy}}{\sigma} + \frac{\alpha^n_{xy}}{\sigma}.
\label{axy}
\end{equation}

An important point is that $\alpha^n_{xy}$ must decrease to zero as $T\rightarrow 0$ because it is a carrier-entropy 
current, just like $\alpha$ (the same is true of $\nu_N$).  By contrast, the vortex-like contribution $\alpha^s_{xy}$ 
strongly diverges as $T\rightarrow T_c$ from above, as the phase stiffness of the superconducting condensate increases.  
If $T_c$ is very small (as in L2), the vortex term appears as a divergent signal as $T$ decreases.  This is a key point in 
what follows. 

\section{Experimental Results}\label{results}
The Nernst experiments are carried out with a thermal gradient (typically 5 K/cm) applied along $\bf\hat{x}$ (the 
longitudinal thermal gradient $-\partial_x T$ is measured with a chromel-alumel thermocouple).  The Nernst voltage 
$E_y$ is measured with a nanovoltmeter (Keithley 2000 with preamp 2001) while the field is slowly ramped at a rate of 
0.4 T/min between -8 to 8 T (or $\pm$ 14 T in high-resolution runs).  Prior to ramping $H$, we regulate $T$ to a 
stability of $\pm$1 mK (this takes 30 min.).  The drift of the voltmeter in zero $H$ ($\sim$10 nV over a 1-hour period) 
is sufficiently small to guarantee a reproducibility in the Nernst coefficient of $\pm$1 nV/KT (about 4 times higher in 
resolution than attained in Ref. \cite{Xu}).  

The need to measure nV signals reproducibly in a temperature gradient introduces specific experimental constraints for 
single crystals.  The present method of sweeping the field (in both directions) at {\em fixed} temperature is quite 
necessary to attain accuracies better than 100 nV/KT.  The alternative, faster, method of sweeping $T$ in a fixed field 
(with a repeat in reversed field) is error-prone and unreliable even if $T$ is swept at the rate of 1 K/min.  The reason is 
apparently the very long relaxation time of the temperature gradient within the crystal.  In our experience, such 
constant-$H$ traces versus $T$, though useful for a qualitative overview of $\nu$, are of limited accuracy.

\subsection{LSCO}
\begin{figure}[htb]
\centerline{\psfig{file=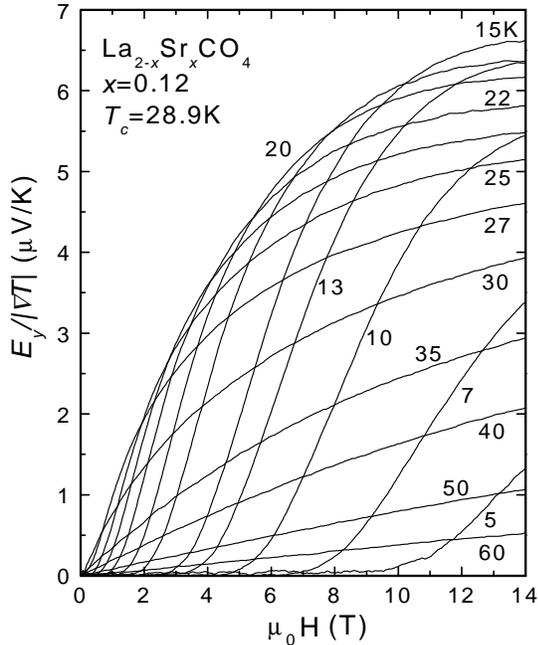,width=2.8in}}
\caption{The field dependence of the Nernst signal $E_y/|\nabla T|$ in $\rm La_{2-x}Sr_xCuO_4$ (Sample L4, 
$x=0.12$) at fixed $T$ from 5 to 60 K.  $E_y$ rises steeply when $H$ exceeds $H_m$ ($\sim$ 5 T at 10 K).  For 
$T$ just below $T_c$ (= 28.9 K), $H_m\simeq 0$, and $E_y$ vs. $H$ shows very pronounced negative curvature.  
The curves continue to display curvature up to 50 K.  Note the relatively slow decay of the signal at 14 T above $T_c$.
}
\label{LSCO}
\end{figure}

\begin{figure}[htb]
\centerline{\psfig{file=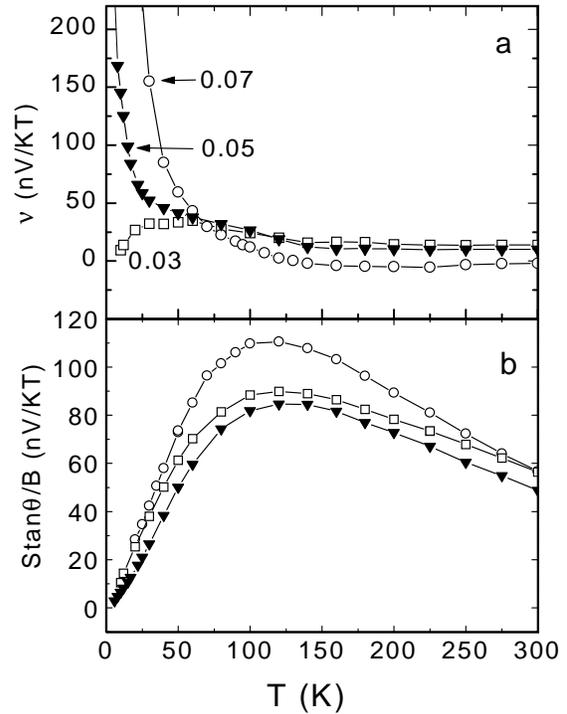,width=3.3in}}
\caption{(a) The $T$ dependence of $\nu$ in very underdoped LSCO (Samples L1, L2, L3, with $x$ = 0.03, 0.05 and 
0.07, respectively).  At low $T$, the different behaviors of $\nu$ in L1 (compared with L2 and L3) indicate distinct 
origins of the Nernst signal. In L1 (open squares), $\nu$ is entirely from the carriers, while in L2 and L3, it is from 
vortex-like excitations.  The lower panel (b) displays $S\tan\theta/B$ measured in L1-L3 (symbol association same as in 
(a)).  Above 150 K, cancellation is nearly complete between the two normal-state currents, $S\tan\theta/B$ and 
$\alpha^n_{xy}$.  At lower $T$, however, a significant residual $\nu_N$ is observed especially in L1.
}
\label{NuLSCO}
\end{figure}

With the higher resolution, we have re-measured $\nu$ in a series of LSCO samples [L1, L2, L3, L4 and L5 with $x$ 
($T_c$) = 0.03 (0 K), 0.05 ($<$ 4 K), 0.07 (12 K), 0.12 (28.9 K), and 0.20 (32.5 K), respectively] ($S$ and 
$\tan\theta$ were also measured in L1-L3, but not in L4 and L5).  The variation of the Nernst signal $E_y/|\nabla T|$ 
with $H$ displayed in Fig. \ref{LSCO} (for Sample L4) is quite representative of LSCO.  At 20 K, for instance, $E_y$ 
rises steeply from zero when $H$ exceeds the melting field $H_m\simeq$ 2 T, approaching saturation at values 6-7 
$\mu$V/K at high fields.  
On warming across $T_c$ (= 28.9 K), the overall magnitude of the signal decreases (as does $H_m$).  Remarkably, 
instead of dropping to zero, the Nernst signal remains large as $T$ rises to 60 K, high above $T_c$.  We note that a 
negative curvature is present even at 40 K.  This would not be expected from the normal-state Nernst signal in a system 
with such short carrier lifetimes.  From such curves, $\nu$ is determined from the initial slope at each selected $T$.  The 
$T$ dependence of $\nu$ is displayed in Fig. \ref{NuLSCO} for the 3 most underdoped LSCO samples.  [For later 
discussion (sec. \ref{pseudo}), we note that $\nu$ continues to increase up to 14 T in the curves at 27 and 30 K.]

As previously reported \cite{Xu}, $\nu$ in Sample L3 ($x= 0.07$) is very small and nearly $T$-independent (-5 to -10 
nV/KT between 300 and 130 K).  Below $\sim$130 K, it begins an inexorable increase that ultimately reaches 2 
$\mu$V/KT at 12 K.  Using the threshold method, we previously identified 130 K as the onset temperature.  However, 
a gradual increase starting near 130 K is also apparent in both Samples L2 ($x$ = 0.05) and L1 ($x$ = 0.03).  In L2, 
$\nu$ continues to increase at lower $T$.  Significantly, however, in the sample with the lowest doping L1, $\nu$ attains 
a broad peak and then {\em decreases} towards zero .  As discussed above, the latter behavior is characteristic of 
$\nu_N$.  

In very underdoped samples, the thermopower is so large ($S\sim 300 \;\mu$V/K at 100 K in L1) that the maximum 
value of $\nu_N$ (40 nV/KT) greatly exceeds the 4 nV/KT threshold.  The juxtaposition of the last 2 curves shows 
especially clearly that the threshold method should be supplanted with another technique more appropriate for samples 
with very large $S$. 

Using the measured thermopower and Hall-angle, we have obtained the curves $S\tan\theta$ displayed in Fig. 
\ref{NuLSCO}(a) for L1-L3.  The 2 panels in Fig. \ref{NuLSCO} illustrates the Sondheimer cancellation at 
temperatures above 150 K.  In L3 ($x = 0.07$), $|\nu|$ is $\sim$10 nV/KT at 150 K whereas $S\tan\theta/B\simeq $ 
110 nV/KT.  Hence, both the normal-state off-diagonal currents $S\tan\theta$ and $\alpha^n_{xy}/\sigma$ must be 
closely matched in magnitude (see Eq. \ref{nuN}).  This cancellation is much less effective at low $T$, especially in L1.

\begin{figure}[htb]
\centerline{\psfig{file=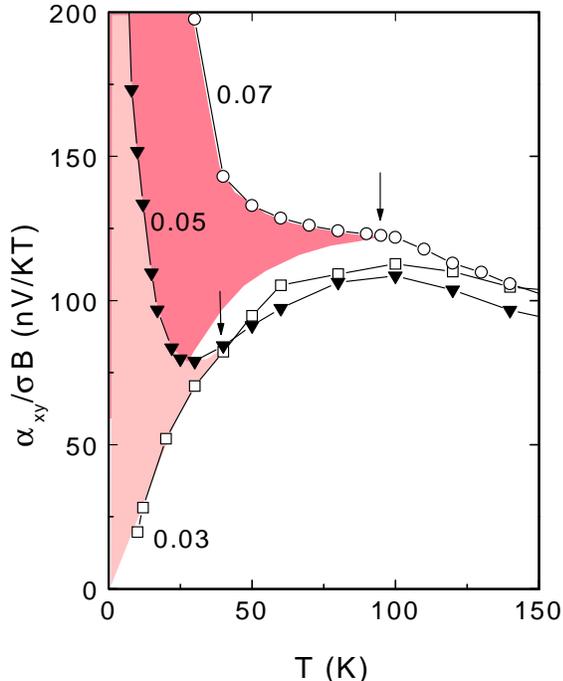,width=3.0in}}
\caption{The $T$ dependence of the Peltier off-diagonal term $\alpha_{xy}$ in L1-L3 obtained by subtracting 
$-S\tan\theta/B$ from $\nu$ (see Eq. \ref{nuall}).  In L1 ($x=0.03$), the profile of $\alpha_{xy}$ (decreasing to zero 
as $T\rightarrow 0$, open squares) identifies it as arising entirely from the carriers.  In L2 ($x=0.05$) and L3 
($x=0.07$), however, the onset of the vortex term $\alpha^s_{xy}$ is apparent as an inflection point (arrows). The 
shaded regions are estimates of the vortex-like term $\alpha^s_{xy}/\sigma B$ in L2 and L3.
}
\label{alphaLSCO}
\end{figure}

Subtracting $S\tan\theta/B$ from $\nu$, we obtain the Peltier curves $\alpha_{xy}/\sigma B$ shown in Fig. 
\ref{alphaLSCO}.  In L1, $\alpha_{xy}/\sigma B$ (open squares) rises to a broad maximum near 100 K before falling 
towards zero as $T\rightarrow$ 0.  This behavior, characteristic of normal carriers, allows us to identify $\alpha_{xy}$ 
with $\alpha^n_{xy}$ at all $T$ in L1.  Sample L2 displays a closely similar profile (solid triangles) except that, below 
50 K, the decrease is interrupted by the onset of the vortex term at 40 K (arrow).  
Hence, removal of the contribution $S\tan\theta$ has made the onset of the vortex term quite unambiguous.  The new 
results show that even at $x=$ 0.05, fluctuation effects arising from pair formation are observed starting at 40 K.  In L3, 
the onset of the vortex-like term at 90 K may also be identified by the inflexion point (arrow).  To estimate the vortex 
contribution $\alpha^s_{xy}/\sigma$ in L2 and L3, we assume their normal contribution $\alpha^n_{xy}/\sigma$ has 
nearly the same form as $\alpha_{xy}/\sigma B$ in L1 apart from a slight re-scaling (as suggested by the similarities in 
the profiles of $S\tan\theta/B$ in Fig. \ref{NuLSCO}(b)).  In L2, we believe this procedure is quite reliable since the 
total $\alpha_{xy}$ is so closely matched to that in L1 above 40 K (and $\alpha^n_{xy}$ must approach zero as 
$T\rightarrow 0$).  In L3, the assumption is less objective.  However, errors incurred (of magnitude $\pm$10 nV/KT) 
are relatively insignificant because $\alpha^s_{xy}/\sigma$ rapidly inflates to values 1-2 $\mu$V/KT.  

The steep increase in the vortex-like contributions in L2 and L3, displayed as the shaded regions in Fig. 
\ref{alphaLSCO},  is rather striking.  In L2, we note that the anomalous signal starts near 40 K even though $T_c$ is 
nominally zero (below  4 K).  These plots represent one of our key findings.

\begin{figure}[htb]
\centerline{\psfig{file=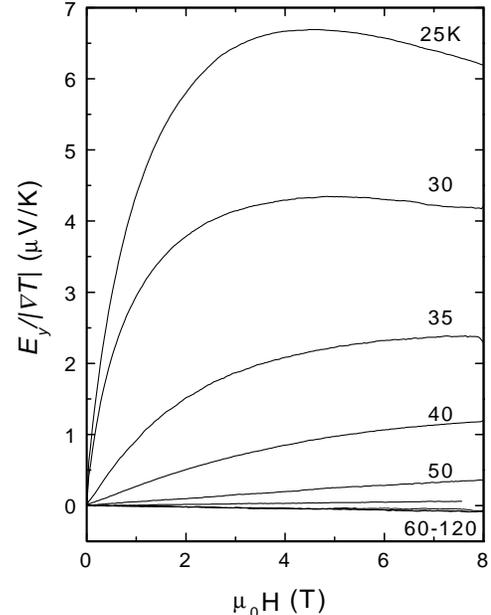,width=2.5in}}
\caption{Variation of $E_y/|\nabla T|$ vs. $H$ in $\rm Bi_2Sr_{2-y}La_yCuO_6$ (Sample B2 with $y$ = 0.5) at 
temperatures 25 to 120 K.  Compared with LSCO, the curves display more pronounced curvature at temperatures high 
above $T_c$. 
}
\label{Bisweep}
\end{figure}

\subsection{Bi 2201}
The systematic variation of transport properties and hole concentration with La content $y$ in single-layer Bi 2201 has 
been investigated by Ando and co-workers \cite{Ando}.  With increasing hole density (decreasing $y$), $T_c$ goes 
through the familiar dome-shaped curve, attaining its maximum value $\sim$35 K at $y$ = 0.4 \cite{Ono}. 

We have investigated 3 samples B1, B2 and B3 in which $y$ ($T_c$) = 0.6 (17 K), 0.5 (29 K) and 0.4 (32 K), 
respectively.  Magnetization measurements show a relatively sharp Meissner transition at $T_c$ and the virtual absence 
of diamagnetic screening above (see below).  The field dependence of the Nernst signal in B2 (Fig. \ref{Bisweep}) is 
representative of the 3 Bi 2201 samples.  Below $T_c$, the trace of $E_y$ vs. $H$ displays highly pronounced 
curvature.  In comparison with LSCO, the negative curvature persists to higher temperatures above $T_c$.  As the hole 
mean-free-path (mfp) is very short (see below), it is clear that the curvature cannot arise from the normal-state 
contribution $\nu_N$.  Rather, the strong sensitivity to these moderate fields is consistent with a superconducting 
fluctuation origin. 

\begin{figure}[htb]
\centerline{\psfig{file=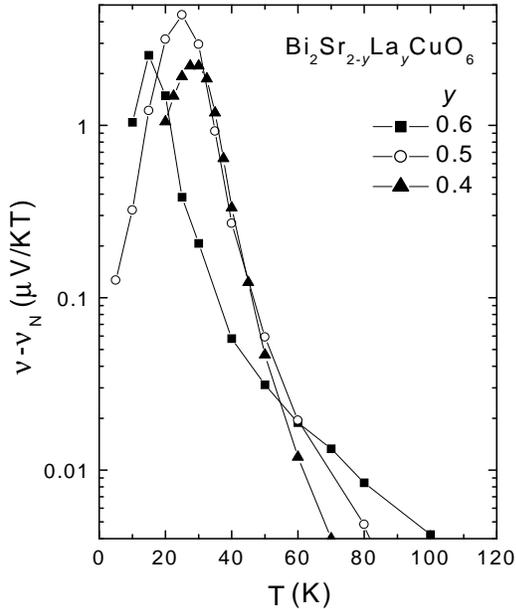,width=2.7in}}
\caption{The $T$ dependence of $\nu$ in Bi 2201 (Samples B1, B2 and B3 with $y$ = 0.6 and 0.5 and 0.4, 
respectively). At all $T$, we determined $\nu$ from the values of $E_y/|\nabla T|$ observed at 1 Tesla (this slightly 
underestimates $\nu$ when curvature is pronounced near $T_c$; see Fig. \ref{Bisweep}).  $T_{\nu}$ decreases 
systematically as the hole concentration increases (from B1 to B3).
}
\label{NuBi}
\end{figure}

The values of $\nu$ measured in a field of 1 T are displayed in Fig. \ref{NuBi}.  The gradual decrease of $\nu$ over an 
extended range of $T$ above $T_c$ bears a striking resemblance to the profiles in LSCO (see Fig. 2 of Xu {\it et al.} 
\cite{Xu}).  The values of $T_{\nu}$ are also quite similar.  

\begin{figure}[htb]
\centerline{\psfig{file=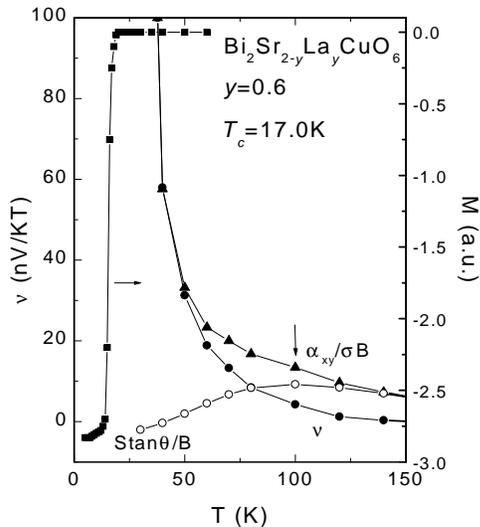,width=2.5in}}
\caption{The $T$ dependence of $\nu$ (solid circles), $S\tan\theta/B$ (open triangles) measured in Sample B1.  The 
Peltier term $\alpha_{xy}/\sigma B$ (solid triangles) is obtained as the sum of these 2 curves (all refer to left scale).  
$T_{\nu}$ is estimated from $\nu$ using the threshold criterion (vertical arrow) ($\nu_N$ is less than 2 nV/KT up to 
300 K). The magnetization curve (from zero field cooling curve measured in a 10 Oe field) shows a sharp Meissner 
transition at $T_c\simeq 17 K$ (solid squares).
}
\label{alphaB1}
\end{figure}
\begin{figure}[htb]
\centerline{\psfig{file=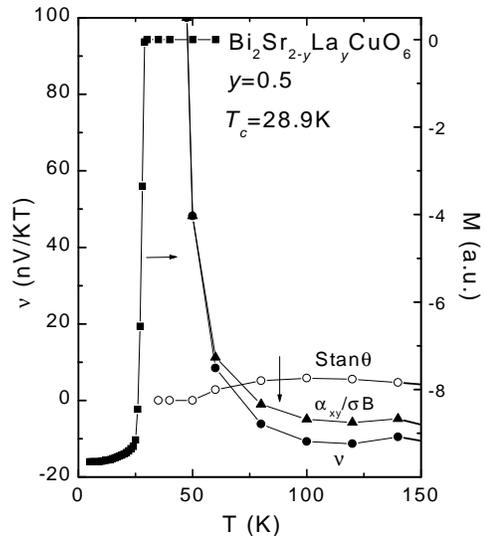,width=2.5in}}
\caption{The $T$ dependence of $\nu$ (solid circles), $S\tan\theta/B$ (open triangles) and $\alpha_{xy}/\sigma B$ 
(solid triangles) in Sample B2 (left scale).  As in Fig. \ref{alphaB1}, $T_{\nu}$ is indicated by the arrow ($\nu_N$ is 
$\sim-10$ nV/KT up to 300 K).  The magnetization curve shows a sharp Meissner transition at $T_c\simeq 29 K$ 
(solid squares).
}
\label{alphaB2}
\end{figure}

We have also measured $S$ and $\tan\theta$ in B1 and B2 (but not in B3).  In contrast to very underdoped LSCO, 
$S\tan\theta/B$ is generally quite small in these samples.  As seen in Figs. \ref{alphaB1} and \ref{alphaB2}, the largest 
values attained by $S\tan\theta/B$, 10 and 6 nV/KT in B1 and B2, respectively, are an order of magnitude smaller than 
in Samples L1-L3.  Most of the suppression comes from the shorter mfp (as determined from $\tan\theta$) in Bi 2201, 
which leads to greatly reduced normal-state off-diagonal currents $S\tan\theta$ and $\alpha^n_{xy}$.

Applying the new method to B1 and B2, we now find that the curves of $\alpha_{xy}/\sigma B$ and $\nu$ differ only 
slightly (by $\sim$10 nV/KT in the range 70-150 K) (Figs.  \ref{alphaB1} and \ref{alphaB2}).  Thus, there is little 
difference (given the measurement uncertainties) whether we use $\nu$ or $\alpha_{xy}/\sigma B$ to estimate 
$T_{\nu}$ (this could not have been anticipated, however, without isolating $\alpha_{xy}$).  The strongly suppressed 
$S\tan\theta$ provides very strong {\em quantitative} arguments against identifying the increase in $\nu$ below 
$T_{\nu}$ with the normal state carriers. [From Fig. \ref{alphaB2}, the observed $\alpha_{xy}$ in B2 increases by 
$\sim$50 nV/KT between 80 and 50 K.  It would be difficult to imagine a scenario in which this increase comes from the 
hole carriers when $S\tan\theta$ is actually decreasing from 6 to $\sim$3 nV/KT over the same temperature range.]  
 
In addition, as mentioned above, the Nernst signal $E_y$ vs. $H$ develops increasingly pronounced negative curvature 
below 50 K.  Curvature in fields less than 8 T cannot arise from hole carriers with such short lifetimes.  We find these 
arguments supporting vortex-like excitations in the range $T_c$ to $\sim$ 80 K especially compelling in Bi 2201.  In 
Figs. \ref{alphaB1} and \ref{alphaB2}, we have also displayed the relatively sharp Meissner transition determined from 
the diamagnetic susceptibility in B1 and B2, respectively.  The comparison emphasizes that in the large interval between 
$T_{\nu}$ and $T_c$, $\nu$ rapidly diverges, but the Meisnner response is essentially non-observable.

\section{Phase Diagram of fluctuations}\label{phase}
\begin{figure}[htb]
\centerline{\psfig{file=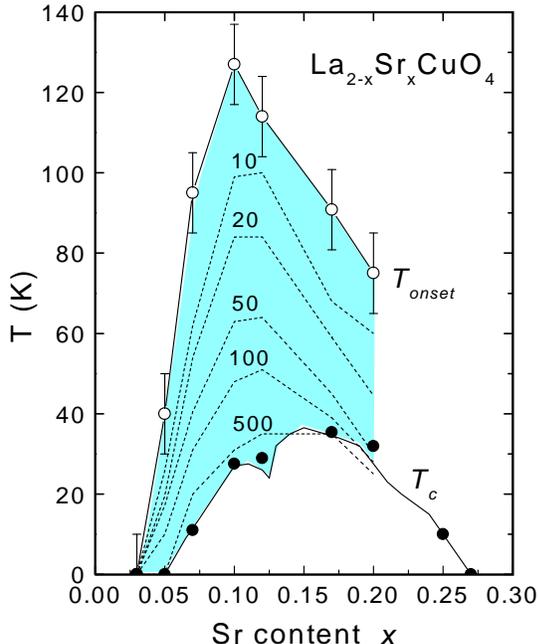,width=2.8in}}
\caption{The $x$ dependence of $T_{\nu}$ and the contours of the vortex-like Nernst signal in LSCO.  For the very 
underdoped samples L1-L3, $T_{\nu}$ is determined from $\alpha_{xy}$ as explained in the text.  The contours are 
those of $\alpha^s_{xy}/\sigma B$ (Fig. \ref{alphaLSCO}).  For samples with $x>$0.07, we have used the procedure 
of Xu {\it et al.} \protect\cite{Xu} to determine $T_{\nu}$ (the contours are the magnitudes of $\nu-\nu_N$).  Data for 
the 2 samples with $x$ = 0.10 and 0.17 are taken from Xu {\it et al.}.  All other data represent new measurements.
}
\label{phaseLSCO}
\end{figure}

With the new results in L1-L3, we have revised our previous phase diagram for LSCO \cite{Xu}.  Figure 
\ref{phaseLSCO} shows the new values of $T_{\nu}$ together with the contours of $\alpha^s_{xy}/\sigma$ estimated 
in Samples L2 and L3 (shaded regions in Fig. \ref{alphaLSCO}).  For samples at higher doping ($x\ge$ 0.10), we 
display the contours corresponding to values of $\nu-\nu_N$, since, as explained above, the new procedure is 
unnecessary if $x>$ 0.07.  

In the revised diagram, vortex-like excitations are absent at all $T$ in the sample with  $x= 0.03$.  Between 0.03 and 
0.07, $T_{\nu}$ increases very steeply from 0 to 90 K with a slope $dT_{\nu}/dx$ of at least 2,400 K (the slope is 
larger if $T_{\nu}\simeq$ 0 at $x=0.04$ as well).  $T_{\nu}$ peaks at $\sim$128 K at $x = 0.11$, and then decreases 
nearly linearly with $x$, but at the slower rate ($dT_{\nu}/dx \sim$ -510 K). 
 
\begin{figure}[htb]
\centerline{\psfig{file=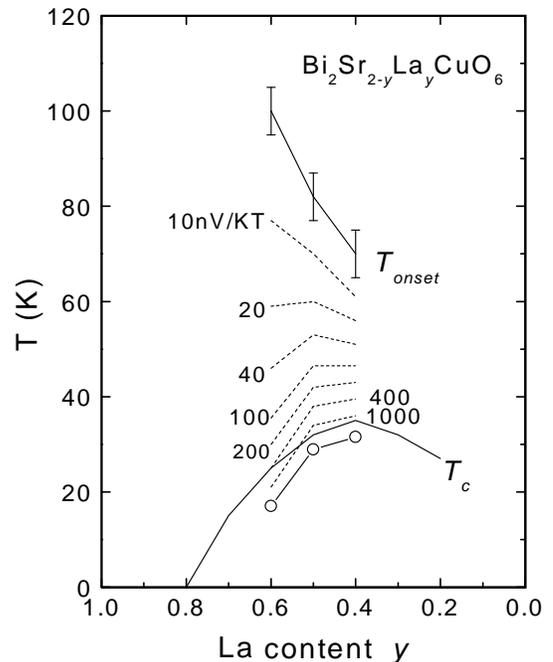,width=2.8in}}
\caption{The $y$ dependence of $T_{\nu}$ and the contours of the vortex-like Nernst signal in $\rm 
Bi_2Sr_{2-y}La_yCuO_6$ determined from $\nu-\nu_N$ using the procedure of Xu {\it et al.} \protect\cite{Xu}.  The 
solid line for $T_c$ is from Ono {\it et al.} \protect\cite{Ono} (open circles are $T_c$ measured in B1-B3).  The 
contours bear a strong resemblance to those in LSCO (Fig. \ref{phaseLSCO}). 
}
\label{phaseBi}
\end{figure}

We have also plotted $T_{\nu}$ and the contours of the anomalous Nernst signal $\nu-\nu_N$ for B1, B2 and B3 in the 
phase diagram of Bi 2201 (Fig. \ref{phaseBi}). $T_{\nu}$ decreases with increasing hole content (decreasing $y$), as 
observed in LSCO for $x>$0.12.  The scale of $T_{\nu}$ is similar to that in LSCO (as is $T_c$).

The phase diagrams in these single-layer cuprates brings out several interesting features.  The fluctuation regime extends 
to a maximum temperature $100-130$ K that is considerably higher than the maximum $T_c$ in either cuprate.  In this 
large interval, vortex-like excitations are readily detected, but no Meissner signal appears until $T_c$ is reached.  The 
contrast is especially dramatic in B1 and B2 (Figs. \ref{alphaB1} and \ref{alphaB2}).  

The onset temperature has a different dependence on doping than $T_c$.  Instead of mimicking the shape of $T_c$ vs. 
$x$, $T_{\nu}$ peaks at 0.11 in LSCO.  Hence, the fluctuation regime is noticeably skewed towards the underdoped 
side (it also extends more deeply into the small-$x$ region than the superconducting phase).  We note that the 
skewedness is evident in all the contours up to 100 nV/KT in Fig. \ref{phaseLSCO}, not just in $T_{\nu}$.  In Bi 2201, 
the less complete data set also show that the fluctuation regime extends to higher $T$ at lower hole concentration (larger 
$y$).  The asymmetry suggests that the strength of the fluctuation regime, as measured by the magnitude of $\nu-\nu_N$, 
tends to increase monotonically with decreasing $x$ (until it suddenly collapses when $x$ is too small).
This general trend suggests that the energy scale associated with pair formation is largest at small $x$, falling nearly 
linearly as $x$ increases.  Because of strong fluctuations, long-range phase coherence -- necessary to sustain a Meissner 
effect -- occurs at a $T_c$ that is much lower than $T_{\nu}$.  

\section{Pseudogap and strong fluctuations} \label{pseudo} 
A key point of interest is the relation of the vortex-like excitations to the pseudogap state \cite{Timusk, Tallon}.  In 
LSCO, information on $T^*$  is largely derived from heat capacity.  The shoulder in $\gamma(T) = c_e(T)/T$ (where 
$c_e$ is the electroinc heat capacity) provides the estimate that $T^*$ equals $\sim$350 K at $x=0.08$ and decreases 
linearly to $\sim$90 K at $x=0.22$ \cite{Tallon,Loram}.  However, there appears to be a lack of corroboration for the 
estimated $T^*$ from nuclear magnetic resonance (NMR) relaxation, infra-red reflectivity, in-plane resistivity $\rho$ or 
tunneling.  The NMR relaxation rate $1/T_1T$ does not show the broad maximum that is a key signature of the 
pseudogap, while $\rho$ does not display the shoulder seen in other underdoped cuprates.

In Bi 2201, Fischer's group recently reported that the gap in the density-of-states observed by scanning-probe tunneling 
spectroscopy \cite{Fischer} is observed up to $\sim$68 K (but the sharp `coherence' peaks at the gap maxima collapse 
at $T_c\simeq$11 K).  The $c$-axis resistivity profile, however, suggests a much higher $T^*$ (100 K, 125 K, and 
250 K for La content $y$ = 0.4, 0.5, and 0.6, respectively) \cite{AndoT}.

The available evidence implies that $T_{\nu}$ is roughly a factor of 2 lower than $T^*$ in LSCO and Bi 2201 (or much 
closer, if we use the tunneling data in Bi 2201).  With the improved resolution for $T_{\nu}$ achieved here, the 
uncertainties now lie chiefly in $T^*$.  [We note, however, that estimates of $T^*$ in any cuprate vary considerably, 
depending on the particular experimental probe used \cite{Timusk}.  The differences may arise from the highly 
anisotropic nature of the pseudogap magnitude and the fact that a particular experimental technique is sensitive to 
excitations at a particular wavevector $\bf q$.]  

With these caveats stated, we note that the contours in Figs. \ref{phaseLSCO} and \ref{phaseBi} are consistent with the 
$x$ dependence of $T^*$ (away from the lightly doped regime).  This suggests that the energy scale for destroying the 
vortex excitations depends on $x$ in the same way as $T^*$.  This trend, suggestive of strong pairing tendencies in the 
lightly-doped regime, recalls early theories of cuprate superconductivity \cite{Anderson} which propose that, in the 
single-layer parent material, pairing correlations are already `built-in' in the spin-$\frac12$ antiferromagnet.  

\emph{Fluctuation regime}    The physical picture suggested is that the pseudogap state, while distinct from $d$-wave 
superconductivity, is nevertheless closely similar in important aspects.  Pairing correlations seem to be already intrinsic at 
high temperatures in the pseudogap state, and fluctuations between the two states become ever stronger as we cool 
away from $T^*$.  Even at temperatures 50-100 K above $T_c$, these pairing correlations begin to support 
vortex-like excitations that are detectable as an anomalous Nernst signal.  Closer to $T_c$ (within 10-20 K), the 
phase-rigidity length inflates dramatically to reach macroscopic length-scales.  This increase is reflected in the rapid 
growth of $\nu$ as reported by Xu {\it et al.} \cite{Xu} (the concomitant increase in kinetic inductance also becomes 
observable at Terahertz frequencies \cite{Corson} in this interval).  In the conventional picture of strong phase 
fluctuations, the case for an extended fluctuation regime above $T_c$ has been made by several theorists 
\cite{Emery,fluc}.  In theories of strongly interacting systems, a discussion of fluctuations between the pseudogap state 
and $d$-wave superconductivity in the SU(2) theory has been published recently by Lee and Wen \cite{Lee}.   

Perhaps the most interesting question raised by these experiments is whether the vortex-like excitations are the familiar 
vortices in a superconducting condensate or novel electronic excitations specific to the pseudogap state.  Is it possible 
that, closer to $T_{\nu}$, the excitations are more properly regarded as vortex-like defects of the pseudogap state 
rather than Abrikosov vortices?  We elaborate further on two points mentioned above. The first is the {\em continuity} 
between the anomalous Nernst signal above $T_c$ and that in the Abrikosov state below.  If we view the Nernst signal 
$E_y/|\nabla T|$ as a function of both $H$ and $T$, it is apparent that the fluctuation regime covers a very wide region 
in the $H$-$T$ plane.  In this paper, we have traced the large fluctuating regime as we move along the $T$ axis in zero 
field.  The fluctuation regime uncovered is actually part of a very large region of the $H$-$T$ plane (as may be seen by 
scanning $H$ at fixed $T$).  As noted in the curves in Fig. \ref{LSCO}, the vortex Nernst signal at temperatures near 
$T_c$ continues to increase with $H$ (up to 14 T) .  We do not observe a decrease of the Nernst signal that might flag 
the crossing of an `$H_{c2}$' line regardless of how close we get to $T_c$.  Detailed analysis of the low-$T$ data 
show that the high temperature fluctuation regime connects continuously with the the high-field fluctuation regime below 
$T_c$ (the extended phase diagram of the fluctuations will be reported elsewhere).  By the continuity argument, the 
vortex-like excitations -- if distinct -- must evolve smoothly into Abrikosov vortices as $T$ decreases towards $T_c$.

The interesting counterpoint is that the onset of the Meissner response is relatively sharp (Figs. \ref{alphaB1} and 
\ref{alphaB2}).  The resistivity profile also implies that conventional amplitude fluctuations in the sense of 
Aslamasov-Larkin \cite{Larkin} occur in a fairly narrow interval.  Unlike in conventional superconductors, the 
diagmagnetic response and resistivity do not `see' the large fluctuation regime uncovered by the Nernst signal above 
$T_c$.  These 2 experimental points are seemingly at odds from the viewpoint of conventional theories of fluctuations, 
but we believe they provide very important hints.

The existence of stable vortex-like defects of the Schwinger-boson condensate in the disordered antiferromagnetic state, 
and their binding to holes, has been discussed by Ng \cite{Ng}.  Resurgent interest in this interesting regime seems likely, 
and we may expect the Nernst effect to play a key role in elucidating its properties (for e.g. its sensitivity to Zn and its 
extension into the overdoped regime).  

{\em Summary}  We have applied the Nernst effect to investigate vortex motion at elevated temperatures in crystals of 
$\rm La_{2-x}Sr_xCuO_4$ and $\rm Bi_2Sr_{2-y}La_yCuO_6$.  To address specifically the lightly doped regime, 
we have adopted a new experimental procedure to sharpen the difference between the vortex-like Nernst signal and the 
carrier Nernst signal near the former's onset temperature $T_{\nu}$.  The total Nernst signal is comprised of a term in 
the off-diagonal Peltier current $\alpha_{xy}$ and a term involving the Hall-angle $S\tan\theta$.  Combining 
measurements of the latter with the Nernst effect, we may back out the $\alpha_{xy}$ current.  We show that in LSCO 
samples with $x\le 0.07$ (in which $S$ is very large), the profile of $\alpha_{xy}/\sigma B$ shows very clearly the onset 
of the vortex-like terms (arrows in Fig. \ref{alphaLSCO}).  The new analysis allows the phase diagram in LSCO to be 
revised to accomodate the lightly doped regime.  The fluctuation regime (which harbors these vortex-like excitations) is 
observed to extend to a maximum temperature of $130$ K and to be skewed towards the underdoped side.  Applying 
the same procedure to 3 samples of La-doped Bi 2201, we show that the carrier contribution to $\nu$ is essentially 
negligible.  Its (partial) phase diagram shares many similar features with that of LSCO.  We emphasize the deep 
penetration of the vortex-like regime into the pseudogap state, and interpret the results in terms of strong fluctuations 
between the pseudogap state and $d$-wave superconducting state.  We discuss whether the vortex-like excitations are 
identical with vortices of the superconducting condensate or novel electronic excitations of the pseudogap state.

\section*{Appendix: Carrier Nernst coefficient in Boltzmann approach}\noindent
We summarize here the standard expressions for the Nernst coefficient in conventional metals. Carriers diffusing in a 
thermal gradient $-\nabla T$ in the presence of $\bf E$ and a weak field $\bf B$ satisfy the Boltzmann equation  
\begin{eqnarray}
{\bf v_k}\cdot \frac{\partial f^0}{\partial \epsilon} 
\frac{(\epsilon_{\bf k} - \mu)}{T}(-\nabla T) 
+ \frac{e{\bf E}}{\hbar}\cdot {\bf v_k} 
\frac{\partial f^0}{\partial \epsilon} \nonumber \\
+ \frac{e{\bf v_k\times B}}{\hbar} \cdot \frac{\partial f_{\bf k}}{\partial {\bf k }} 
= -\frac{g_{\bf k}}{\tau_{\bf k}},
\label{Boltz}
\end{eqnarray}
where $ g_{\bf k} = f_{\bf k}- f^0_{\bf k}$ is the difference between the perturbed distribution function $f_{\bf k}$ 
and its value $f^0_{\bf k}$ at equilibrium ($\bf v_k$ and $\tau_{\bf k}$ are the velocity and lifetime in state $\bf k$).  
To find the Peltier conductivity elements, we may set $\bf E$ = 0, and expand $g_{\bf k} = g^{(0)}_{\bf k} + 
g^{(1)}_{\bf k} + \cdots$, where $g^{(0)}$ (linear in $-\nabla T$) gives $\alpha$ while $g^{(1)}$ (linear in $-\nabla 
T$ and $B$) gives the off-diagonal term $\alpha_{xy}$. By iteration of Eq. \ref{Boltz}, we have
\begin{eqnarray}
g^{(0)}_{\bf k} & = & -\tau_{\bf k} \frac{\partial f^0}{\partial \epsilon}
{\bf v_k}\cdot \frac{(\epsilon_{\bf k} - \mu)}{T} (-\nabla T),\\
g^{(1)}_{\bf k} & = & \frac{e{\bf v_k\times B}}{\hbar} \cdot 
\frac{\partial g^{(0)}_{\bf k}}{\partial {\bf k }}.
\label{g}
\end{eqnarray}
The off-diagonal current $J_y$ is just $eg^{(1)}{\bf v_k}$ integrated over the Fermi Surface.  Hence, the off-diagonal 
conductivity is (in terms of the mean-free-path $\vec{\ell}{\bf (k) \equiv v_k}\tau_{\bf k}$)
\begin{equation}
\alpha_{yx} = e^2\sum_{\bf k}  
\frac{(\epsilon_{\bf k} - \mu)}{T}
\left(-\frac{\partial f^0}{\partial \epsilon}\right) \ell_y 
\frac{{\bf v_k\times B}} {\hbar} \cdot
\frac{\partial \ell_x}{\partial {\bf k}}.
\label{alphaxy}
\end{equation}
In 2D systems (with arbitrary dependence of $\ell({\bf k})$ on $\bf k$), we may use the swept-area representation 
\cite{Ong} to reduce this complicated expression to 

\begin{equation}
\alpha_{xy} = \frac{2e^2B}{(2\pi)^2T\hbar^2} 
\int d\epsilon \left(-\frac{\partial f^0}{\partial \epsilon}\right)
(\epsilon -\mu) {\cal A}_{\ell}(\epsilon),
\label{alpha2D}
\end{equation}
where ${\cal A}_\ell(\epsilon) = \oint d\ell_x \ell_y $ is the area swept out by $\vec{\ell}({\bf k})$ as $\bf k$ goes 
around a contour at energy $\epsilon$.  As $\sigma_{xy}$ is proportional to ${\cal A}_\ell(\mu)$ \cite{Ong}, we find
\begin{equation}
\alpha_{xy} = \frac{\pi^2}{3}\frac{k_B^2T}{e}
\left[\frac{\partial\sigma_{xy}}{\partial\epsilon} \right]_{\mu}.
\label{alphaxy}
\end{equation}
This recalls the familiar relation between $\alpha$ and $\sigma$
\begin{equation} 
\alpha = \frac{\pi^2}{3}\frac{k_B^2T}{e} \left[\frac{\partial \sigma}{\partial \epsilon}\right]_\mu.
\label{alpha}
\end{equation}

Substituting Eqs. \ref{alphaxy} and \ref{alpha} into Eq. \ref{nuN}, and using the small Hall-angle approximation 
($\sigma_{xy}/\sigma\simeq \theta$), we have 
\begin{equation}
\nu_N  =  \left[ \frac{\alpha_{xy}}{\sigma} - \frac{\sigma_{xy}}{\sigma}\frac{\alpha}{\sigma} \right]\frac1B 
   =  \frac{\pi^2}{3}\frac{k_B^2T}{e} \frac{\theta}{B} 
\left[ \frac{\partial \ln\theta}{\partial \epsilon} \right]_\mu.
\label{nuN2}
\end{equation} 
If $\theta$ is only weakly dependent on energy at $\epsilon_F$ (a good approximation in conventional metals), we have 
nearly exact cancellation of the two contributions and $\nu_N$ is very small \cite{Sondheimer}.  

To estimate $\nu_N$ in case $\theta(\epsilon)$ has an anomalously strong power-law dependence $\epsilon^p$, we 
write $\nu_N \simeq (86\;\mu{\rm V/K})[k_BT/\epsilon_F] (\theta/B)p$. Using the values $\epsilon_F\sim$ 0.5 eV, 
and $(\theta/B)\simeq 4 \times 10^{-3}$ at 100 K (to approximate optimum-doped LSCO), we find that $\nu_N 
\simeq 6p$ nV/KT, which is still small.

{\em Isothermal versus adiabatic conditions}  In Nernst experiments on metals in which the electronic thermal 
conductivity $\kappa^e$ dominates the phonon conductivity $\kappa^{ph}$, the (electronic) thermal Hall conductivity 
$\kappa_{xy}$ generates a transverse gradient $-\partial_y T$.  This leads to a transverse current comparable to the 
other transverse currents in Eq. \ref{Jy}, which now reads
\begin{eqnarray}
J_y &= & \alpha_{yx}(-\partial_x T)  + \sigma_{yx}E_x + \alpha(-\partial_y T)+ \sigma E_y\nonumber\\
& = & \left[\alpha_{yx} - \sigma_{yx}\frac{\alpha}{\sigma}\right](-\partial_x T) + \alpha(-\partial_y T) + \sigma E_y = 
0.
\label{Jyk}
\end{eqnarray}
Under adiabatic conditions (the transverse edges are free to `float' to different temperatures), we have $-\partial_y T = 
(\partial_x T)\kappa_{yx}/[\kappa^{ph}+\kappa^e]$.  Assuming $\kappa^{e}\gg\kappa^{ph}$, and writing 
$\kappa_{yx}/\kappa^e \equiv \eta\sigma_{yx}/\sigma$, where $\eta$ is a positive number of order 1, we get the 
equation  
\begin{equation}
\left[\alpha_{yx} - (1+\eta)\sigma_{yx}\frac{\alpha}{\sigma}\right](-\partial_x T) + \sigma E_y = 0.
\end{equation}
Physically, the transverse thermal gradient produces a current that augments the Hall current $\sigma_{yx}E_x$.

Technically, under adiabatic conditions the measured $E_y$ does not give the Nernst coefficient.  Heroic efforts are 
required to `short out' the transverse gradient (for example, by thermally anchoring the transverse edges to each other 
with a piece of thick wire).  Fortunately, in the cuprates, the phonon conductivity $\kappa_{ph}$ is more than 10 times 
larger than $\kappa^e$.  Hence, the phonons act as a shorting fluid that keeps $\partial_y T$ negligibly small.  
Measurements of $\kappa_{xy}$ in YBCO are reported by Zhang {\it et al.} \cite{Zhang}.  In LSCO, $\kappa_{xy}$ 
is much smaller, and barely detectable \cite{Krishana}.  The standard geometry in Fig. \ref{Expt} is effectively in the 
isothermal limit for conductors in which $\kappa^e\ll \kappa^{ph}$.\\
\\
\centerline {* \quad * \quad *}\\

The research at Princeton is supported by the U.S. National Science Foundation (MRSEC grant NSF-DMR 
98-09483).  N.P.O. and S. U. acknowledge support from a grant from the New Energy and Industrial Tech. Develop. 
Org., Japan (NEDO).  We wish to thank P. W. Anderson, J. Clayhold, S. Kivelson, P. A. Lee, A. J. Millis, V. N.  
Muthukumar, J. Orenstein and Z. Weng for helpful dicussions.

\bigskip
$^\dagger${\it Present address of ZAX. Department of Physics, Zhejiang University, Hangzhou 310027, China.}

\end{document}